\documentclass[aps,preprint]{revtex4}
\usepackage{amsmath}
\usepackage{graphicx}
\usepackage{amssymb}

\begin{document}

\preprint{BROWN-HET-1450}

\title{Statistical entropy of two-dimensional dilaton de Sitter space }

\author{David A. Lowe}

\email{lowe@brown.edu}

\affiliation{Physics Department, Brown University, Providence, RI 02912, USA}

\begin{abstract}
It has been proposed that a quantum group structure underlies de Sitter/Conformal
field theory duality. These ideas are used to give a microscopic operator
counting interpretation for the entropy of two-dimensional dilaton
de Sitter space. This agrees with the Bekenstein-Hawking entropy up
to a factor of order unity. 
\end{abstract}
\maketitle

\section{Introduction}

String perturbation theory is defined using the notion of a finite
number of string loops moving in a fixed background spacetime. This
modest starting point yields a finite perturbative expansion for scattering
amplitudes. It has had remarkable success in elucidating many non-perturbative
phenomena, such as strong/weak coupling duality symmetries, the existence
of new solitonic objects, D-branes, as well as the microscopic interpretation
of black hole entropy in certain cases. Moreover these new ideas have
led to the first conjectures for complete (though background dependent)
non-perturbative formulations of string theory, such as Matrix theory
and AdS/CFT.

However it seems clear that we are still missing some key ideas. In
the present paper, we will be mainly concerned with developing the
idea that noncommutative geometry and quantum groups should play a
much more prominent role than it has to date. This idea has already
been extensively studied for both closed (see \cite{Frohlich:1995mr}
for a review) and open strings in compact backgrounds (see for example
\cite{Alekseev:2000wg}). It has long been known that WZNW models
based on the compact group $g$ have an underlying quantum-deformed
symmetry $U_{q}(g)$ where the deformation parameter $q$ is a root
of unity, determined by the level number of the CFT. There are a finite
number of unitary irreducible representations of the quantum group
which correspond to integrable representations of the current algebra.
The tensor product structure of the quantum group determines CFT operator
product coefficients. If one then attempts to reconstruct the target
space geometry by Fourier transforming this finite set of representations,
one winds up with a non-commutative space on which the quantum group
symmetry acts. For the case of D-branes on group manifolds, the non-commutative
geometry that emerges makes contact with ideas of Connes \cite{A.Connes1994}. 

The hope is that quantum group structure may build in from the beginning
analogs of spacetime uncertainty principles (see \cite{Li:1998fc}
for a review), and thus provide a formulation of string theory applicable
to the strong gravity regime where it no longer makes sense to think
of single strings moving in smooth spacetime background. In order
to make these ideas more precise, it is interesting then to further
consider the examples of AdS/CFT and dS/CFT \cite{Strominger:2001pn}
and look for quantum group symmetry underlying these dualities. This
was first studied in \cite{Jevicki:1998rr} for the case of AdS/CFT
where it was found that the quantum group symmetry of WZNW models
gives a natural explanation of the stringy exclusion principle \cite{Maldacena:1998bw}
for the case of $AdS_{3}\times S_{3}$. Related ideas appear in \cite{Steinacker:1998kv,Steinacker:1999xu,Ho:1999bn,Jevicki:2000ty}.
From the spacetime viewpoint, this bound is explained by giant gravitons
\cite{McGreevy:2000cw}, whose maximum size is cutoff by the radius
of the sphere.

It was proposed in \cite{Guijosa:2003ze} that similar ideas could
be extended to dS/CFT by seeking an underlying quantum group symmetry
as a $q$-deformation of the isometry group of de Sitter/conformal
group of the CFT. An important new feature of this construction is
the appearance of non-compact groups. It was shown that cyclic unitary
representations of the quantum group $U_{q}(SL(2))$ become unitary
principal series representations in a classical limit. These are precisely
the representations corresponding to massive particle states in a
two-dimensional de Sitter background. For $q$ a nontrivial root of
unity the quantum group representations are finite dimensional, so
the spectrum of the theory becomes discrete. Thus the quantum group
structure introduces both an ultraviolet and an infrared cutoff in
an interesting way. These are prerequisites for a microscopic interpretation
of the finite horizon entropy of de Sitter. Generalization to three-dimensional
de Sitter was discussed in \cite{Lowe:2004nw} and a formula for the
de Sitter entropy was proposed. This was further described from the
classical dS/CFT viewpoint in \cite{Guijosa:2005qi}. Related ideas
and further developments may be found in \cite{Pouliot:2003vt,Banks:2005bm}.

In the present paper these ideas are applied to the case of two-dimensional
dilaton de Sitter space \cite{Cadoni:2002kz,Medved:2002tq}. This
spacetime has a nontrivial Bekenstein-Hawking entropy and temperature.
The entropy of this spacetime is accounted for by counting operators
in a dual $q$-CFT.

\section{Two-dimensional dilaton de Sitter space}

\begin{figure}
\includegraphics{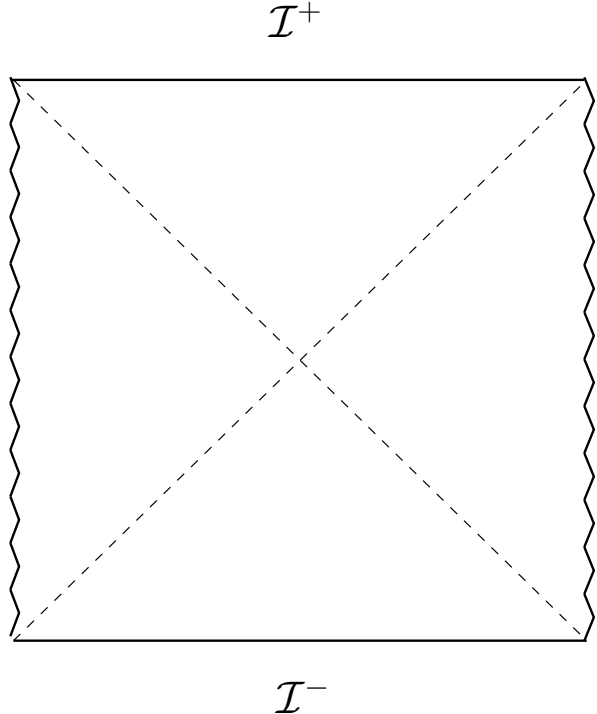}

\caption{\label{cap:Penrose-diagram}Penrose diagram for de Sitter-like solution
of two-dimensional dilaton gravity. There is a singularity at one
point on the spatial circle where the dilaton $\Phi=0$ and the gravitational
coupling diverges. This is analogous to the conical singularity of
three-dimensional Schwarzschild de Sitter.}
\end{figure}

As described in \cite{Cadoni:2002kz} two-dimensional dilaton gravity\[
S=\frac{1}{2}\int\sqrt{-g}d^{2}x~\Phi\left(R-\frac{2}{\ell^{2}}\right)\]
 admits solutions of de Sitter form\begin{eqnarray}
ds^{2} & = & -\frac{1}{\frac{t^{2}}{\ell^{2}}-a^{2}}dt^{2}+\left(t^{2}-a^{2}\ell^{2}\right)d\theta^{2}\nonumber \\
\Phi & = & \Phi_{0}\frac{t}{\ell}~,\label{eq:dssol}\end{eqnarray}
where $a$ is a dimensionless constant parameterizing a mass deformation,
$\Phi_{0}$ is a dimensionless constant parameterizing the strength
of the gravitational constant, $\ell$ is a length scale that sets
the radius of curvature. We take $\theta\in(0,2\pi)$ to parametrize
a spacelike circle and $t\in(-\infty,0)$. The Penrose diagram for
this solution is shown in figure \ref{cap:Penrose-diagram}. The $(t,\theta)$
coordinates only cover the lower quadrant of the diagram.

This spacetime has an asymptotic symmetry group with a Virasoro algebra
that acts \cite{Cadoni:2002kz,Medved:2002tq}. As described in \cite{Guijosa:2005qi},
states of matter fields in this background fall into representations
of this Virasoro algebra.%

There exists definition of mass $M$ utilizing a spacelike Killing
vector on $\mathcal{I}^{-}$ \cite{Cadoni:2002kz}. The mass defined
in this way can be positive or negative. On the solution (\ref{eq:dssol})
\begin{equation}
|M|=\frac{a^{2}}{2\ell}\Phi_{0}~.\label{eq:cmass}\end{equation}
In the limit $a\to0$ the singularity becomes null, and the geometry
is that of a big crunch or big bang. The $a=1$ limit describes a
pure de Sitter geometry. We expect the region $0\leq a\leq1$ to correspond
to sensible semiclassical spacetimes. This can be argued by lifting
to three dimensions where these geometries correspond to a positive
deficit angle. On the other hand, the $a>1$ geometries will correspond
to negative deficit angle. Since this would arise from a negative
energy localized source, we do not expect these geometries to be stable
once interactions are included. It is curious the allowed geometries
have $|M|$ less than the de Sitter value. However $M$ measures the
energy of the complete spacetime, and addition of positive energy
sources causes the cosmological horizon to shrink, reducing the overall
contribution to $|M|$.

Demanding absence of a singularity on the Euclidean section leads
to a Hawking temperature\begin{equation}
T=\frac{a}{2\pi\ell}~.\label{eq:thawk}\end{equation}
Because the dilaton varies, the relation of Bekenstein-Hawking entropy
to area is a bit more subtle than the standard case. The entropy may
be derived by computing the Lorentzian action of the solution with
appropriate boundary terms included. However the simplest way to obtain
the entropy is to take $1/T=\partial S/\partial M$ which leads to\begin{equation}
S=2\pi\Phi_{0}a\label{eq:bekhawk}\end{equation}
where the constant of integration is fixed so $S=0$ when $T=0$.
This prescription agrees with results obtained via the semi-classical
Lorentzian path integral.

We can try to infer something about the spectrum of microscopic excitations
by asking at what temperature do we expect the thermodynamic limit
to break down. This happens when $S\sim\mathcal{O}(1)$ so $a\sim1/\Phi_{0}$.
Therefore an estimate of the mass gap above the $T=0$ solution is\begin{equation}
\Delta M\sim\frac{1}{\Phi_{0}\ell}~.\label{eq:levspac}\end{equation}

\section{$q$-deformed CFT}

The proposal for describing two-dimensional $dS_{2}$ using a $q$-deformed
version of the dS/CFT correspondence has been studied in \cite{Guijosa:2003ze}.
The interpretation of horizon entropy in this framework has been elaborated
in \cite{Lowe:2004nw,Guijosa:2005qi}. 

Based on those results we postulate the $q$-deformed CFT is built
out of $N$ representations of the $q$-deformed isometry group $SL(2,\mathbb{R})$
characterized by the parameters $\tau=-1$ and complex number $b$
as described in detail in appendix \ref{sec:Complementary-Series}.
We set $q=e^{2\pi i/N}$ a root of unity. The $q$-deformed representations
are analogs of the complementary series representations for massless
particles in de Sitter space (strictly speaking mass $\to0^{+}$,
and recall the relation $\tau=-\frac{1}{2}-\sqrt{{\frac{1}{4}-(m\ell)^{2}}}$
when $m\ell<1/2$). 

We have in mind that the full interacting CFT will be based on a theory
with $SU(N)$ gauge symmetry, or something similar, but we will presume
there exists a free-field limit where one is left with $N$ fields.
On this branch we will assume the gauge symmetry is broken down to
the permutation group $S_{N}$ with the $N$ fields transforming in
the defining representation. This is reminiscent of the structure
of the CFT's relevant for the D-brane black holes of \cite{Strominger:1996sh}.

The results of \cite{Lowe:2004nw} carry over for the spectrum of
these representations. The generator $L_{0}$ is associated with the
Killing vector used to define the notion of mass (\ref{eq:cmass})
(here we follow the conventions of \cite{Guijosa:2003ze,Lowe:2004nw},
which differ from \cite{Guijosa:2005qi,Cadoni:2002kz}). This is related
to the generators defined in appendix \ref{sec:Complementary-Series}
by the relation $L_{0}=(X_{+}+X_{-})/2$. In particular, for one of
these $N$-dimensional cyclic representations with $b$ real, $L_{0}$
has imaginary eigenvalues that are roughly equally spaced ranging
from $-i(N-1)/2\ell,\cdots,i(N-1)/2\ell$. In this way we see the
$q$-deformation introduces a ultraviolet and an infrared cutoff versus
the continuous unbounded spectrum of the $q=1$ principal series representation.
Note also the spacing of the $L_{0}$ eigenvalues does not approach
a continuum in the $q\to1$ limit. 

Related issues arise in the fat-black hole limit when one considers
black holes with D-brane charge \cite{Maldacena:1996ds}. In that
case the mass gap one would infer by having $N$ distinct D-strings
wrapping a one-cycle is much larger than the lowest inverse length
scale in the system. There the resolution is that the $N$ distinct
D-strings coalesce into a single multiply wrapped string. This picture
produces the expected mass gap.

An analogous phenomena can happen for the cyclic representations.
We assume the dynamics is such that instead of all $N$ representations
having the same value for $b$ that they differ by a phase $e^{2\pi ik/N}$
where $k=1,\cdots,N$ with $k$ labelling the $N$ different cyclic
representations. This possibility was explored in \cite{Lowe:2004nw}
for the case of the principal series. These results also carry over
to the complementary series described here. The end result is one
obtains a reducible representation with a spectrum for $L_{0}$ ranging
from approximately $-i(N-1)/2\ell,\cdots,i(N-1)/2\ell$ with a typical
level spacing of order $1/N\ell$. Therefore we match the expected
spacing (\ref{eq:levspac}) provided we identify $N$ with the strength
of the gravitational coupling\begin{equation}
N\sim\Phi_{0}~.\label{eq:nplanck}\end{equation}
 This gives the expected continuous spectrum for $L_{0}$ as $N\to\infty$.

Now we are ready to compute the de Sitter entropy arising from this
microscopic description, following \cite{Lowe:2004nw,Guijosa:2005qi}.
A comoving observer in de Sitter sees a thermal density matrix due
to a trace over modes outside their horizon. In \cite{Lowe:2004nw,Guijosa:2005qi}
this notion was carried over to the CFT description. In particular,
the partition function becomes a sum over a operators in the CFT with
Boltzmann weights. The operators that appear in the sum are those
corresponding to the one-particle modes with positive imaginary $L_{0}$
eigenvalue, together with their tensor products. We begin by assuming
Bose statistics for these modes. The expectation value of the mass
then collapses to a sum over single particle modes with the Bose-Einstein
distribution function appearing\[
M=\sum\frac{L_{0}}{e^{L_{0}/T}-1}\approx N\ell\int_{0}^{(N-1)/2}dL_{0}\frac{L_{0}}{e^{L_{0}/T}-1}\sim\ell NT^{2}\]
for $N\gg1$, $T\ell\gg1/N$, and where we have identified $T$ with
the Hawking temperature. 

Plugging in to get the entropy we find\[
S\sim\ell NT~.\]
Therefore, recalling equations (\ref{eq:thawk}), (\ref{eq:bekhawk})
and (\ref{eq:nplanck}) we find agreement with the Bekenstein-Hawking
entropy up to a factor of order unity. In the limit $N\gg1$, replacing
the Bose-Einstein distribution by Fermi-Dirac changes the overall
coefficient by a constant factor of order unity.

\section{Discussion}

In this paper we have provided a statistical computation of entropy
in a de Sitter-like spacetime. We computed the microscopic entropy
of a two-dimensional version of de Sitter space that depends on two
non-trivial parameters: a gravitational constant and a mass parameter.
In the usual semi-classical picture, this entropy diverges, but by
$q$-deforming the spacetime this entropy is rendered finite and can
be given a dual interpretation as an operator counting problem in
a $q$-deformed CFT. 

Clearly much remains to be done in further developing the correspondence
between $q$-deformed CFT's and quantum de Sitter spacetimes. Thus
far it has been established that unitary CFT's of this type exist,
and that these appear to account for the entropy of de Sitter space
in a straightforward way. It would be very interesting to find versions
of string theory that live in these backgrounds and higher dimensional
generalizations, and to give a complete specification of the dual
boundary CFT's. 

To conclude, let us offer the following speculation. Suppose a CFT
of the type outlined in the present work can be shown to be a complete
self-consistent theory dual to a gravitational theory in asymptotic
de Sitter space in four spacetime dimensions. This presumably will
belong to a family of different theories, labelled by the relevant
gauge symmetry (lets say $SU(N))$. From the bulk point of view, this
will correspond to a family of disconnected string vacua. While its
possible we started out in some special state in the past (for example
special states with much larger effective cosmological constants leading
to inflation), as time evolves we expect to evolve into the most likely
type of microstate accessible to us. Therefore if we rule out the
unstable spacetimes analogous to the $a>1$ backgrounds considered
here, the CFT description predicts this will be the macrostate with
the largest available entropy, which is de Sitter with a small positive
cosmological constant determined by $N$. This matches well with current
observations and leads to the prediction that dark energy density
will asymptote to a constant determined by the fundamental constants
of the theory.

\begin{acknowledgments}
I thank A. G\"uijosa for helpful comments. The research of D.A.L.
is supported in part by DOE grant DE-FE0291ER40688-Task A and NSF
U.S.-Mexico Cooperative Research grant \#0334379. 
\end{acknowledgments}
\appendix

\section{Complementary Series\label{sec:Complementary-Series}}

In \cite{Guijosa:2003ze} a $q$-deformed version of the principal
series representations of $SL(2,\mathbb{R})$ was found. Here we generalize
those results to the case of the complementary series. Let us begin
by reviewing the complementary series for the case $q=1.$ We can
realize this representation on the basis $|k\rangle=e^{-ik\theta}$
with $k$ an integer. The action of the generators takes the form\begin{eqnarray*}
He^{-ik\theta} & = & 2ke^{-ik\theta}\\
X^{+}e^{-ik\theta} & = & (k-\tau)e^{-i(k+1)\theta}\\
X^{-}e^{-ik\theta} & = & -(k+\tau)e^{-i(k-1)\theta}\end{eqnarray*}
where for the complementary series $-1<\tau<0$. As described in \cite{Klimyk:1991fg}
there is an equivalence of representations under $\tau\to-1-\bar{\tau}$.
We will also be interested in the discrete series representation corresponding
to $\tau=-1$. For our purposes, this may be considered a continuation
of the complementary series.

The complementary series is unitary with respect to the norm\begin{equation}
\langle\chi|\psi\rangle=\sum_{n=-\infty}^{\infty}\frac{\Gamma(\tau-n+1)}{\Gamma(-\tau-n)}a_{n}\bar{b}_{n}\label{eq:cnorm}\end{equation}
where\[
\psi=\sum_{n=-\infty}^{\infty}a_{n}e^{-in\theta}~,\qquad\chi=\sum_{n=-\infty}^{\infty}b_{n}e^{-in\theta}~.\]
The coefficients appearing in the norm (\ref{eq:cnorm}) arise when
the Klein-Gordon norm is computed for a field in de Sitter with mass
$0\leq m<(d-1)/2$, as shown in appendix \ref{sec:Klein-Gordon-norm}.

The $q$-deformed version of the algebra takes the form\begin{equation}
KK^{-1}=K^{-1}K=1,\qquad KX^{\pm}K^{-1}=q^{\pm2}X^{\pm},\qquad[X^{+},X^{-}]=\frac{K-K^{-1}}{q-q^{-1}}~.\label{eq:qalg}\end{equation}
where the classical limit is obtained by setting $K=q^{H}$ and taking
the limit $q\to1$. We will be interested in the case where $q=e^{2\pi i/N}$
with $N$ odd. The basic structure of the representations of interest
can be carried over from the results of \cite{Guijosa:2003ze}, 

\begin{eqnarray*}
K|m\rangle & = & q^{-2m}\lambda|m\rangle\\
X^{+}|m\rangle & = & \left(bc+\frac{q^{m}-q^{-m}}{q-q^{-1}}\frac{\lambda q^{1-m}-\lambda^{-1}q^{m-1}}{q-q^{-1}}\right)|m-1\rangle\\
X^{-}|m\rangle & = & |m+1\rangle\end{eqnarray*}
with $m=0,\cdots,N-1$, $\lambda=q^{2l}$, $l=(N-1)/2$ and $b$,
$c$ complex numbers that satisfy

\begin{equation}
bc=\tau^{2}+\tau-l^{2}-l~.\label{eq:abeqn}\end{equation}
These transformations are supplemented by the cyclic operations\[
X_{+}|0\rangle=b|N-1\rangle~,\qquad X_{-}|N-1\rangle=c|0\rangle~.\]
To check unitarity, we need to check positivity of the norm (\ref{eq:cnorm}).
The eigenvalues of $H$ are real, since $l$ is an integer. It remains
to examine\begin{eqnarray}
\left\langle X^{+}m|X^{+}m\right\rangle  & = & -\left\langle m|X^{-}X^{+}|m\right\rangle \nonumber \\
 & = & -\left(bc+\frac{q^{m}-q^{-m}}{q-q^{-1}}\frac{\lambda q^{1-m}-\lambda^{-1}q^{m-1}}{q-q^{-1}}\right)\langle m|m\rangle\label{eq:unorm}\end{eqnarray}
if we use the notion of conjugation defined by the {*}-structure $X_{\pm}^{*}=-X_{\mp}$,
$K^{*}=K^{-1}$. Substituting in for $bc$ , we need to check whether\[
v=l^{2}+l-\tau(\tau+1)-\left(\frac{q^{m}-q^{-m}}{q-q^{-1}}\frac{\lambda q^{1-m}-\lambda^{-1}q^{m-1}}{q-q^{-1}}\right)>0\]
This can be expressed as \[
v=l^{2}+l-\tau(\tau+1)-\frac{\sin\left(\frac{2\pi(l-k)}{2l+1}\right)\sin\left(\frac{2\pi(l+1+k)}{2l+1}\right)}{\sin^{2}\left(\frac{2\pi}{2l+1}\right)}=l^{2}+l-\tau(\tau+1)+\frac{\sin^{2}\left(\frac{2\pi(l-k)}{2l+1}\right)}{\sin^{2}\left(\frac{2\pi}{2l+1}\right)}\]
Since $-1\leq\tau\leq0$ this expression is always positive, so the
$q$-deformed representation is unitary.

\section{Klein-Gordon norm\label{sec:Klein-Gordon-norm}}

In this appendix we demonstrate the coefficients appearing in the
norm (\ref{eq:cnorm}) arise from the Klein-Gordon norm when $0\leq m\ell\leq1/2$.
The Klein-Gordon norm is

\begin{equation}
(\psi,\phi)=-i\int d\Sigma^{\mu}(\psi\overleftrightarrow{\partial_{\mu}}\phi^{*})~.\label{eq:kknorm}\end{equation}
 Rather than working with the coordinate patch (\ref{eq:dssol}),
we will instead work in global coordinates \[
ds^{2}=-dT^{2}+\ell^{2}\cosh^{2}T~d\theta^{2}\]
to make direct contact with previous work \cite{Guijosa:2003ze,Lowe:2004nw}
and take $\theta$ to have $2\pi$ periodicity. The dilaton vanishes
at $\theta=0$ in these coordinates. In the mass range of interest
the mode expansion for a free minimally coupled scalar field takes
the form\[
\phi=\sum_{n=-\infty}^{\infty}f_{n}(T)e^{-in\theta}\]
where $\phi$ satisfies the equation\[
\square\phi=m^{2}\phi\]
The solutions take the form\[
f_{n}=A_{n}P_{-\frac{1}{2}-n}^{-\tau-\frac{1}{2}}\left(\tanh\frac{T}{\ell}\right)+B_{n}Q_{-\frac{1}{2}-n}^{-\tau-\frac{1}{2}}\left(\tanh\frac{T}{\ell}\right)\]
where $\tau=-\frac{1}{2}-\sqrt{\frac{1}{4}-(m\ell)^{2}}$ and the
coefficients are chosen so the modes are orthonormal with respect
to the Klein-Gordon norm (\ref{eq:kknorm}). This imposes the condition\[
A_{n}^{*}B_{n}-A_{n}B_{n}^{*}=\frac{i}{\pi^{5/2}}\frac{\Gamma(\tau-n+1)}{\Gamma(-\tau-n)}\]
The $n$ dependent factor in this expression reproduces the nontrivial
$n$ dependent factor that appear in the norms of the complementary
series (\ref{eq:cnorm}).

\bibliographystyle{/home/lowe/current/brownphys}
\bibliography{qdscft}

\end{document}